\documentclass{article}
\usepackage{graphicx} 
\usepackage[utf8]{inputenc}
\usepackage{biblatex}
\usepackage{indentfirst}
\usepackage{array}
\usepackage{booktabs}
\usepackage{multirow}
\usepackage{caption}
\usepackage[edges]{forest}
\usepackage{tikz}
\usepackage{adjustbox}
\usetikzlibrary{shadows.blur}
\usepackage{caption}

\usepackage{xurl} 
\usepackage{hyperref} 

\title{The Epistemological Consequences of Large Language Models: Rethinking collective intelligence and institutional knowledge}
\author{Angjelin Hila}
\date{November 2024}

\begin{document}

\maketitle

\begin{abstract}
In this paper we interrogate the epistemological implications of human-LLM interaction with a specific focus on epistemological threats. We develop a theory of epistemic justification that synthesizes internalist and externalist conceptions of epistemic warrant termed collective epistemology. Collective epistemology considers the way epistemological warrant is distributed across human collectives. In pursuing this line of thinking, we take bounded rationality and dual process theory as background assumptions in our analysis of collective epistemology as a mechanism of collective rationality. Following this approach, we distinguish between internalist justification as a robust standard of rationality and externalist justification as a reliable knowledge transmission mechanism. We argue that while these standards jointly constitute necessary and sufficient conditions for collective rationality, only internalist justification produces knowledge. We posit that reflective knowledge entails three necessary and sufficient conditions: a) rational agents reflectively understand the basis on which a proposition is evaluated as true b) in absence of a reflective evaluative basis for a proposition rational agents consistently evaluate the reliability of truth-sources, and c) rational agents have an epistemic duty to apply a) and b) as rational standards in their domains of competence. Since distributed rationality is socially scaffolded, we pursue the consequences of unchecked human-LLM interaction on social epistemic chains of dependence. We argue that LLMs approximate a type of externalist justification termed reliabilism but do not instantiate internalist standards of justification. Specifically, we argue that LLMs do not possess reflective justification for the information they produce but rather reliably transmit information whose reflective basis has been established in advance. Since LLMs cannot produce knowledge with reflective justifiedness but only reliabilist justifiedness, we argue that human outsourcing of reflective knowledge to reliable LLM information threatens to erode reflective standards of justification at scale. As a result, LLM information reliability disincentivizes comprehension and understanding in human agents. Human agents that forfeit comprehension and understanding for reliably correct results reduce the net justifiedness of their own beliefs and, consequently, reduce their ability to perform their epistemic duties professionally and civically. The scaled outsourcing of reflective knowledge to LLMs across collectives entails an impoverishment of the production of reflective knowledge. In order to mitigate these potential threats, we propose the development and promotion of epistemic norms across three tiers of social organization: a) normative epistemic model for individual human-LLM interaction, b) norm setting through institutional and organizational frameworks optimal epistemic outcomes and c) the imposition of deontic constraints at organizational and/or legislative levels in order to instill LLM discursive norms that reduce epistemic vices.

\end{abstract}

\section{Introduction}
In this paper we consider the epistemological consequences of human-LLM interaction both at the individual and collective scale. We develop a theory of reflective knowledge that builds on the existing philosophical literature from modern epistemology. Briefly, we take \textit{reflective knowledge} to be the joint possession of true beliefs by an epistemic agent and the justification of those true beliefs through access to well-grounded reasons. Extrapolating to collective rationality, we define \textit{collective justification} as the distribution of epistemic justification across communities of epistemic agents. We argue that collective justification necessitates normative epistemological standards that balance reflective justification as the highest standard of knowledge and reliabilist justification as the standard of knowledge trustworthiness. Subsequently, we examine how the growing integration of human-LLM interaction into personal and institutional workflows impacts reflective epistemic norms. We propose a three-tiered framework to mitigate the erosion of justified knowledge, encompassing: (1) individual guidelines for human-LLM interaction, (2) institutional normative frameworks, and (3) deontic or legislative constraints to curb harmful effects. Accordingly, our paper makes contributions to information science, philosophical epistemology, and science and technology studies (STS). Our information science contributions are twofold: a) tracing the effects of integrated human-LLM interaction on institutional information flows and b) integrating these epistemological consequences with state-of-the-art models in information seeking and retrieval. Our contributions to philosophical epistemology involve advancing a novel theory of collective justification that integrates internalist and externalist standards of justification and extending virtue epistemology to human-LLM interaction contexts. Our contributions to STS involve identifying concrete threats posed by state-of-the-art AI technology on social processes of collective knowledge production and transmission and proposing a mitigatory framework with a view to promote virtuous human-LLM interaction practices that preserve and promote the reflective epistemic standards upon which human collective action depends. In order to draw the epistemological implications of human-LLM interaction, we first set the emergence of LLMs in historical context. By tracing the development of LLMs to the early history of computation and the inception of the AI research program in the 1950s, we set the arena for understanding human-LLM interaction in the context of automation, philosophical theories of computation and mind, and the unique threats that LLMs pose in contrast to prior computational paradigms.

\section{From Computers to AI}

Two approaches have dominated the field of artificial intelligence since its modern inception in the 1950s Dartmouth College: symbolic AI and connectionism. Symbolic AI attempts to model artificial intelligence by analogy to “thought” or symbol manipulation through formal rules. By contrast, connectionism attempts to model artificial intelligence by analogy to the intrinsic structure of biological brains, which comprise networks of cells called neurons that communicate through electro-chemical signals \cite{smith2019promise}. In the Symbolic AI paradigm, the chief mechanism of learning is logical inference. In the connectionist paradigm the mechanisms of learning consist of signal propagation through distributed parallel processing (DPP) in a network of neurons. 

In order to better understand these two models of intelligence, it helps to give a brief overview of the historical development of computation. The modern artificial intelligence research programme has its roots in the emergence of computers in the 1940s. The electrical computer draws on three strands of groundbreaking developments: a) the development of mathematical logic in the 19th century, b) the development of electrical boards in the early 20th century and c) the theoretical foundations of computer science established by Alan Turing, Claud Shannon and John von Neumann (stored program concept) in the 1930s \cite{nisan2005elements, petzold1999code}. 

\subsection{Foundations of Computation}

Modern mathematical logic begins with George Boole’s publication of \textit{An Investigation of the Laws of Thought in 1847}, in which he developed a logical system that expressed statements of logic in terms of algebraic classes and sets. Boole expressed logical disjunction and conjunction in terms of the mathematical operations of addition and multiplication\cite{boole2022investigation}. This groundbreaking equivalence between logic and algebra demonstrated that logical operations could suffice in engineering a general computational machine. It wasn’t until the development of electricity, however, that the successful marriage between physical machine and algorithmic abstraction came to fruition. In 1936, Alan Turing published \textit{On Computable Numbers, With An Application to the Entscheidungsproblem}, wherein he developed a mathematical model of a universal computing machine \cite{turing2004computable}. A year later, Claude Shannon mathematically demonstrated in his Master’s Thesis, \textit{A Symbolic Relay of Switching Circuits}, that electrical circuits could implement bivalent Boolean algebra \cite{shannon1938symbolic}. Together, these two formal innovations constitute the genesis of modern computer science and the basis for the general purpose computer. 

Turing’s innovation involved specifying a minimal logical model that carries out any computable operation or algorithm called a Universal Turing Machine. The Universal Turing Machine constitutes an abstract blueprint for the space of possible computational machines. In other words, the Turing Machine delimits the set of all logically possible computable functions with implications for the physical implementation of such a machine \cite{turing2004computable}. In principle, a Turing Machine can carry out any computable function, while in practice a universal computer faces physical limitations of memory and storage that bar it from realizing all possible Turing computable algorithms \cite{turing2004computable}. In computer science, programming languages that in principle can realize any algorithm computable by a Turing Machine bear the designation Turing-Complete. 
	
Shannon’s implementation of logical operations through electrical circuitry demonstrated how a Turing Machine could be constructed as a physical mechanism. For example, the AND operation can be implemented through series-connected relays where a closed switch represents 1 and an open switch represents 0 \cite{shannon1938symbolic}. With these two serial inputs the logical truth values for AND can be realized where two closed switches represent the value \textit{true} and the other permutations the value \textit{false} \cite{shannon1938symbolic}. Conversely, the OR operation can be implemented through parallel-connected relays where two open parallel switches represent the value \textit{false}, and the rest of the permutations represent the value \textit{true} just as in the logic table for the disjunction operation \cite{shannon1938symbolic}. 

These theoretical foundations that led to the emergence of modern computers spawned both the symbolic and connectionist approaches toward general artificial intelligence. The former approach identified software as the hallmark of intelligence and therefore sought to model intelligence on rule-based symbol manipulation. The latter approach, on the other hand, placed equal emphasis on the hardware architecture that implements a program or algorithm and aimed to model intelligence on the network structure of the brain. Since the advent of computers involved the implementation of logic within a physical circuit, it provided the impetus for symbolic AI and a variety of theories of the human mind termed computational theories of mind \cite{fodor1975language}. Symbolic AI and the computational theories of mind share the assumption that intelligence reduces to formal symbol manipulation \cite{smolensky1987connectionist}. The computational theorists expressed this relationship with the slogan: “the brain is to the hardware as the mind is to the software” \cite{searle2006chinese}, suggesting that mental states are analogous to computer software. Given this assumption, it is natural that early attempts at artificial intelligence sought to model intelligence on symbolic reasoning. The so-called Expert Systems that dominated the artificial intelligence market in the 1980s and 1990s emerged as the representative paradigms of the Symbolic AI era \cite{jackson1990expert}. 

\subsection{Expert Systems}

Expert systems integrate four components: a knowledge base, an inference engine, a knowledge acquisition component, and a user interface. The knowledge base hosts the domain-specific facts and domain-specific reasoning rules. Knowledge representation techniques such as semantic networks are deployed for representing hierarchical relationships between objects \cite{lucas1991principles, tzafestas1993overview}. The inference engine defines the sequential steps for selecting and executing the appropriate inference rules on the knowledge base. It administers the rules in a cyclical manner in order to extract new information and terminates when no further inferences can be generated. The rules generally consist of conditional statements such as “if the profits fall below x threshold, then decrease production by y amount” \cite{jackson1990expert}. The knowledge acquisition component stores solved scenarios into its knowledge base, thereby expanding the system’s knowledge store. Users interact with the system through the user interface by either feeding it facts or asking it how to solve a particular problem. When the user prompts the system to reason from available knowledge to some general conclusion, the inference engine utilizes a technique called forward chaining \cite{lucas1991principles}. On the other hand, when the user prompts the system to begin with the conclusion and reason its way back in order to provide the necessary evidence, the inference engine deploys a technique called backward chaining \cite{lucas1991principles}. Finally, the expert system justifies the inference to the user through an explanatory module. 
	
While expert systems systematize human expertise, they exemplified the failure of the Symbolic AI paradigm to model general intelligence \cite{gill1995early}. Where a human expert may err either in their application of reason or consideration of evidence, an expert system applies inference rules consistently and utilizes the available evidence to generate sound conclusions and optimal decisions. However, because expert systems rely on human designers to supply both the data and the inference rules, which require in-depth and domain-specific knowledge in order to carry out their intended functions, they are not capable of autonomously learning patterns from data like human subjects \cite{duchessi1995understanding}. Because they lack flexible learning mechanisms and adaptive behaviours exhibited by intelligent systems, they fail to model fundamental aspects of general intelligence that enable it to update its models from new data and became synonymous instead with highly-specialized reasoning machines. 

\subsection{Connectionism}

The roots of the connectionist paradigm also trace back to the early days of artificial intelligence with researchers like McColloch, Pitts, and Rosenblatt seeking to implement a Turing Machine through a network architecture of artificial neurons. McCulloch and Pitts hypothesized that the excitatory and inhibitory action potentials of biological neurons could be implemented to carry out Boolean functions \cite{mcculloch1943logical}. An action-potential is a quasi-binary electro-chemical signal that biological neurons transmit when sufficient depolarization occurs between positively charged sodium ions outside the neuron membrane and negatively charged potassium ions in the neuron body. Patterns of excitation and inhibition distributed across vast networks account for information processing and transmission, although the exact mechanisms the brain employs to realize mental functions elude present neuroscience. 

Improving upon McColloch and Pitt’s artificial binary input-and-output neuron model called the Threshold Logic Unit, Rosenblatt constructed the first perceptron, a single-layer artificial neural network that takes any real-valued inputs and reduces the error of the output by computing the difference from the expected mean value \cite{rosenblatt1958perceptron}. Minksy and Papert introduced an activation function to Rosenblatt’s perceptron that better simulated the biological action-potential by converting the output into a continuous nonlinear function \cite{minsky1969perceptrons}. Nonetheless, the single-layer perceptron computationally implements only a simple linear classifier and famously cannot simulate some logical functions like the exclusive or (XOR) that require more than a single linear split \cite{minsky1969perceptrons}. While Minsky and Papert recognized the potential of multilayer perceptrons for rectifying this limitation and enabling universal function approximation, they forewent investigation into multilayer perceptrons. Subsequently, two AI winters ensued, as they came to be known, first with the perceived failure of the perceptron to simulate the XOR (exclusive or) function required for the functional completeness of a computational system, leading to a funding draught for artificial neural networks research, and second, during the late nineteen eighties and early nineteen nineties due to unmet expectations of the revolutionary potential of AI to usher in a new era of automation \cite{muthukrishnan2020brief}.

The thawing of the AI winter was precipitated by the resurgence of the backpropagation algorithm in the mid to late 1980s \cite{muthukrishnan2020brief}. One of the architectural challenges with artificial neural networks concerned improving the performance of the network. The network computes the output by propagating the inputs through the hidden layers and outputting a numerical prediction in the output layer \cite{lecun2015deep}. Each neuron takes as input the sum of products of the values of the outputs of each neuron in the previous layer and the weights associated with each connection plus a scalar value for the bias \cite{lecun2015deep, babcock2021generative}. The input neuron passes this sum of products plus the bias through an activation function that serves as a threshold for passing the output to the next layer. The value of the weights and biases determine whether the activation function propagates the output to the next layer \cite{lecun2015deep, babcock2021generative}. Designers employ differing initialization strategies for the weights and biases such as randomizing them for the first iteration \cite{babcock2021generative}. The network outputs are converted into probability ranges between zero and one before being fed into the loss function that computes the network error as a function of the expected values or ground truth of the prediction \cite{lecun2015deep, babcock2021generative}. Once the error is computed, the challenge lies in changing the network parameters in order to minimize the error. Prior to the advent of backpropagation, the weights and biases were updated through a number of techniques, including manually, heuristic rules or through evolutionary algorithms where the best performing parameters among a population of networks were selected \cite{babcock2021generative}. 

The innovation of backpropagation involved a method of algorithmically computing the gradient of the entire neural network error by retroactively computing the gradient of each layer in order to adjust the weights and biases \cite{babcock2021generative}. The gradient refers to a vector-valued function that describes the composite function of the entire network (or individual layers). Computing the gradient of the error involves differentiating the vector function through the chain rule of calculus to find the derivative of the gradient \cite{lecun2015deep}. The sign of this derivative indicates the direction of highest ascent, namely the direction of the local maximum. Since the objective is to minimize the error globally, an optimization strategy involves taking a step in the opposite direction of the gradient, namely the direction of the local minimum, with the aim of finding the global minimum \cite{lecun2015deep}. Once the optimization step is executed for the network error, backpropagation executes a backward pass for each of the hidden layers, computing the error for each of the neurons \cite{lecun2015deep}. Computing the error for each layer allows the optimization algorithm to adjust the weights and biases accordingly by either increasing or decreasing their values toward maximal error minimization \cite{lecun2015deep, babcock2021generative}. Forward and backward passes are iterated until the error converges on the lowest possible value or global minimum \cite{lecun2015deep}. 

The application of backpropagation to multilayer neural networks was first proposed by Paul Werbos in his PhD thesis (1974), where he argued that propagation of the error signal from the output back to the input through gradient descent could be used for weight optimization. However, it wasn’t until “Learning representations by back-propagating errors” by Rumelhart, Hinton and Williams 
\cite{rumelhart1986learning} that significant improvements were proposed that made backpropagation efficient. Rumelhart et al. demonstrated that backpropagation in a multi-layer perceptron could be used to learn hierarchical representations and extract features directly from data \cite{rumelhart1986learning}.  

\subsection{LLMs}

As artificial neural networks began to garner more attention due to performance improvements, two architectures dominated the landscape: convolutional neural networks in image and vector processing and recurrent neural networks in natural language processing. Convolutional neural networks employ convolutional layers to first extract high-level features from an image and pooling layers to reduce the dimensionality of the extraction \cite{lecun2015deep}. The convolutional layer consists of a two or three dimensional matrix of weights called a kernel that is multiplied to the input tensor in order to extract relevant features. Both the convolutional and pooling layers involve matrix multiplication with a bitmap representation of the input in n dimensions \cite{lecun2015deep}. Convolutional networks employ a hierarchy of combinations of convolutional and pooling layers in order to capture global features of images such as object detection \cite{lecun2015deep}. The initial convolutional layer detects lower level features such as edges, textures and simple shapes, while subsequent layers detect mid to higher level features such as combinations of edges and patterns all the way up to whole objects. Various architectures and strategies for optimal image processing exist such as AlexNet, which employs five convolutional layers and 3 max pooling layers \cite{lecun2015deep}. 
	
Recurrent neural networks (RNNs) and subsequent variants such as Long Short-Term Memory (LSTM) networks employ feedback and a type of memory called a hidden state in order to store the previous output, which is subsequently fed as input to the next iteration \cite{schmidhuber1997long, vanhoudt2020review, graves2012long}. The use of feedback enables RNNs to remember ordered sequences of data, a necessary step for the network to preserve word relationships across textual sequences. In a basic recurrent network the second output is generated by the input as well as the previous output stored as an n-dimensional vector called the hidden state \cite{medsker1999recurrent, grossberg2013recurrent}. The dimensionality of the hidden state vectors determines the capacity of the network for contextual processing by representing the sequence of inputs up to the current step \cite{grossberg2013recurrent}. This strategy faces problems of high dimensionality that lead to exploding and vanishing gradients \cite{babcock2021generative}. Vanishing gradients refer to gradients of complex functions where the variation is so minimal that the network cannot learn by reducing the error \cite{bengio2016deep}. Exploding gradients, conversely, refer to extremely large gradient ranges that prevent further optimization \cite{bengio2016deep}. Both of these outcomes limit the capacity of the network for representing long-term dependencies between words. 

Long-Short-Term-Memory networks (LSTM) address this memory bottleneck by employing a three-gate structure consisting of a forget, input, and output gates that allows the network to determine the relevance of subsequent inputs for retention \cite{schmidhuber1997long}. The LSTM unit employs an additional vector called a cell state to regulate outputs of the hidden state \cite{vanhoudt2020review, graves2012long}. The LSTM unit uses the forget gate to weight inputs against previous states in order to decide whether to forget or retain them in effect ensuring that the cell state gradients do not degrade over time \cite{staudemeyer2019understanding}. The input gate determines what new information should be added to the cell state and the output gate determines the information that should be fed as output to the next hidden state \cite{staudemeyer2019understanding, yu2019review}. In other words, the cell state regulates long-term dependencies as input and output to the hidden state that feeds previous outputs as inputs to the next state. While these improvements to the architecture enable LSTM networks to learn word and meaning dependencies across longer sequences of textual data, they still process textual information sequentially and rely on encoding dependencies within a single vector, which limits the context window to broadly local dependencies and hinders the network from learning longer-term dependencies that describe global properties of a textual corpus \cite{graves2012long}. 

The innovation of the Transformer architecture availed NLP architectures from their intrinsic bottleneck problem by innovating a mechanism called self-attention that enables parallel distributed processing of textual information. Like RNN and LSTM architectures, the Transformer architecture makes use of a word representation method called word embeddings. Word embeddings convert each word into a dense vector representation, meaning a vector with typically more than two real-numbered values, that represents the location of that word relative to a fixed vocabulary in n-dimensional space \cite{rumelhart1986learning, mikolov2013distributed}. Locational closeness in the n-dimensional space captures similarities between words. The greater the number of dimensions the more nuanced the representation of word similarities. However, unlike RNNs and LSTMs, the self-attention mechanism processes textual data strings in a parallel fashion by converting the input into query-key-value vectors \cite{vaswani2017attention, cheng2016long}. When a sentence is fed into the architecture, the Transformer treats each word as a query and compares it to every word in the sentence (including itself) through a similarity measure \cite{babcock2021generative}. Each word is represented by a key that accesses its value, namely the vector representation in the embedding. This similarity measure outputs a score that is normalized into a [0,1] ranged weight that is then multiplied to each word value and summed to produce a new vector representation of each word that captures its relational context or “meaning” in the input string \cite{vaswani2017attention}. This mechanism creates a representation of word-to-word relevance in the string. By simultaneously treating all words as queries, self-attention computes the contextual relevance of each word to all the other words, in effect encoding the meaning of a sentence, paragraph or larger body of text without requiring sequential processing. However, since Transformers eschew sequential processing they also lose the word order encoding that is essential for processing natural languages \cite{babcock2021generative}. To recover word order Transformers incorporate positional encodings, a method that includes the sequential position of a word in its vector representation \cite{babcock2021generative}. 

\subsection{Institutional Ramifications}

Our survey of the two paradigms of intelligence, symbol manipulation and connectionism, provides a blueprint for understanding the interaction of evolving information processing systems with information institutions such as libraries. As organs of collective intelligence, libraries service collective intentionality through the application, maintenance, and transmission of collective endeavours such as the law, engineering, medicine, information technologies, and cultural practices. Advances in artificial intelligence stand to reconfigure the processes and products underlying collective intentionality. While the interaction vector between artificially intelligent systems and domains of human competence is difficult to predict, we identify the potential that LLMs carry to disrupt information chains through wide-scale personal and institutional adoption. We argue that indiscriminate adoption without necessary normative controls can compound the erosion of individual knowledge and a fortiori the epistemic controls that organizations require to function properly. 

\section{Paradigms of Automation}

In 2016, Klaus Schwab coined the phrase the Fourth Industrial Revolution to denote the advances marked by a cluster of interrelated and intersecting technologies as ushering a new industrial phase from the preceding three periods of industrial change \cite{schwab2017fourth}. These encompass advances in artificial intelligence and its effects on industrial design and production, the interconnectedness of devices and systems dubbed the internet of things, and the increasing integration of biological and mechanical engineering due to advances in gene editing and nanotechnology \cite{schwab2017fourth}. With respect to preceding revolutions, the first consisted of the shift from manual labour to steam and mechanical power, the second involved the erection of railways and telegraph systems as revolutionizing transportation and communication, and the third phase took place toward the middle of the 20th century with the advent of electronic computers and new levels of automation that it ushered. 

Analogously, the philosopher of information Luciano Floridi has coined the neologism the Fourth Revolution to designate our current information age with respect to the cultural evolutionary lineage of information technologies \cite{floridi2014fourth}. According to Floridi, besides our own, three revolutionary stages in information technology have shaped the trajectory of human civilization. These include the Neolithic or agricultural evolution when humanity transitioned from prehistory into history by developing recording technologies that enable cultural transmission, the industrial revolution in the late seventeenth and early eighteenth centuries, and the information revolution marked by radio and telegraph transmission \cite{floridi2014fourth}. In contrast to preceding information-technological revolutions, Floridi refers to our age as hyperhistory  \cite{floridi2014fourth}. 

The recent success of generative AI marks an inflection within our hyperhistorical age because it engenders a mode of cultural and information production that qualitatively differs from previous machine learning capabilities. Generative AI empowers artificial intelligence with generative and creative capabilities comparable to and perhaps surpassing those of humans. The degree to which the outputs of Generative AI amount to mere ersatz reproductions of genuine human creativity and ingenuity is currently an open question. Defaulting to the null hypothesis that Generative AI merely resuscitates what human ingenuity has already wrought, the quality and speed with which it produces facsimiles and recombinations of text and audiovisual content marks an unprecedented shift in the methods of cultural production. The alternative hypothesis that Generative AI produces novel outputs at least in some domains of content generation additionally complicates their status as automation technologies. In light of these capabilities, whole regions of human labour stand to be hybridized by Generative AI, markedly shifting or redistributing creative agency between human and AI agents. 

Even though the Transformer language processing architecture that underlies LLMs was developed in 2017 \cite{vaswani2017attention}, the release of ChatGPT in 2022 showcased to the public at large nearly human-indistinguishable natural language proficiency in chatbots. The implications of this technology for a wide range of professional and organizational domains remain to be seen. Chatbots can generate 200-300 words per minute of scientific-level content. Compositional natural language generation through LLMs automates a once uniquely human area of competence. Until the emergence of LLMs, humans were the only species who could produce compositional language, which enables humans to produce knowledge and coordinate mass collective action. The degree to which LLMs wield language with the referential and causal richness of humans is a matter for debate. The author thinks that LLMs cannot understand language, which, in tandem with consciousness and thought, uniquely endows humans with reflective knowledge. Until AI replicates reflective knowledge, humans will remain the epistemic standard for verifying ground-truth. However, the possession of reflective knowledge notwithstanding, the accuracy, reliability, and output speeds of LLMs stand to transform and augment a whole range of professional competences. This redistribution of agency showcases weaknesses in frameworks like actor-network theory, which posit the parity between human and non-human agencies \cite{latour2007reassembling}. Currently, human plans subsume technological agency. However, the possible distribution of professional labour to Generative AI will engender new human-out-of-the-loop systems that altogether eschew human oversight and decision-making \cite{mellamphy2021humans}. A society where information systems are managed and maintained by autonomous and adaptive artificial intelligent systems becomes easier to envisage. Having situated LLMs within the lineage of industrial production and automation, we turn to their epistemological properties. I argue that evaluating the epistemic status and powers of LLMs carries implications for rational human agency and particularly institutional trust. 

\section{The Epistemology of AI Chatbots}

We draw from the core epistemological theories to analyze whether AI Chatbots meet rational standards for knowledge. We argue that while AI Chatbots increasingly meet reliabilist criteria for knowledge, they fail to meet internalist criteria for knowledge. We defend an internalist view of epistemology as the hallmark of human knowledge, namely the rational standard upon which all knowledge-transmission relies. 

Epistemology stands for the theory of knowledge. The orthodox theory of knowledge defines knowledge minimally as justified true belief \cite{morton2008guide}. In this definition, the conditions of truth, belief, and justification are understood to be logically independent from one another. Rational agents may possess strong grounds for believing propositions that turn out to be false, or they may hold beliefs that turn out to be true by accident such as guessing. Since justified belief does not guarantee truth and true belief can be accidental eliding rational justification, modern theory of epistemology treats justification as a fallible but truth-conducive condition aimed at maximizing knowledge. In other words, modern theory of knowledge seeks to ascertain the conditions that render the connection between belief and truth rational. As such, philosophers ask what justifies an agent’s beliefs \cite{morton2008guide}. To this effect, theories of justification diverge along the internalist-externalist axis \cite{morton2008guide}. The predominant models of internalism are foundationalism and coherentism. Foundationalism holds that an agent's beliefs must be inferentially reducible to basic, non-inferential beliefs in order to be justified. Coherentism holds that an agent's beliefs must form an inferentially coherent network in order to attain justification. Externalism, by contrast, rejects the requirement that justification depend on the agent’s internal access, instead treating justification as any reliably truth-conducive process, independent of the agent’s awareness of its reliability (Morton 2008). In addition, a great body of philosophical literature explores whether justified true belief constitute necessary and sufficient conditions for knowledge. The famous Gettier cases argue that justified true beliefs are necessary but insufficient conditions for knowledge \cite{gettier2020justified}. The discussion of Gettier cases falls beyond the scope of this article. We focus on theories of justification as endowing agents with rational grounds for belief and externalist theories of knowledge, which deny that access to the rational basis of belief is necessary for knowledge.

In addition to orthodox epistemology which takes knowledge to be minimally justified true belief, we take Herbert Simon’s notion of bounded rationality as a constraining factor on the net justifiedness of an agent’s web of belief. Simon defined bounded rationality as decision making under uncertainty or lack of perfect information \cite{simon1996sciences}. In light of uncertainty, bounded rational agents opt for satisficing or sufficiently good rather than rationally optimal decisions \cite{simon1996sciences}. We argue that bounded rationality, namely the assumption that agents operate with limited information and under conditions of uncertainty, implies collective rationality: agents undertake rationally circumscribed tasks and outsource related rational tasks to other rational agents forming a dependent chain or network where rational action ranges over the entire chain. This does not mean that individual agents operate as rationally optimal agents, but rather that they are tasked with applying the highest rational standards available to them in the domains in which they exercise rational responsibility. This entails a normative model of rationality where its application requires adherence to truth-conducive epistemic norms: e.g. good reasoning, accumulation of relevant evidence, seeking parsimonious explanations, and mitigating cognitive biases. 

Since instrumental rationality, namely the pursuit of outcomes, depends on the justifiedness and truthfulness of agents’ beliefs, collective rationality implies collective epistemology: agents outsource belief-formation to reliable processes within the wider collective. Simon’s construal of rationality as bounded dovetails with the great rationality debate unfolding in the last few decades in the fields of psychology and cognitive science. Stanovich distinguishes between weak rationality as conceived in the Aristotelian, categorical sense, where rationality categorically includes the human species and excludes the rest, with strong rationality as a continuous normative model which agents can approximate only suboptimally \cite{stanovich2010rationality}.  We take the latter model of rationality as a trait that agents exhibit in degrees in our analysis of the epistemological effects of human-LLM interaction. In contrast to Stanovich, however, we are not concerned with rationality as a personal trait aimed to maximizing self-interest, but theoretical rationality as a set of truth-conducive practices. Further, we are concerned with rationality as a subpersonal trait that construes agents as embedded in systems and social norms that inculcate and sanction varying levels of rational behaviour. 

\subsection{Internalism}

Internalism encompasses a family of epistemological theories that locate the standard of knowledge in the relation of the knower to the bases of their beliefs. Species of internalism have been developed in psychology and cognitive sciences in response to growing evidence on cognitive biases and heuristics \cite{kahneman2012thinking, stanovich2010rationality, johnson1983mental}. Following the broad distinction within dual-process theory between automatic processing and reflective and explicit processing, we will defend an internalist account of justification that we will call reflective justifiedness. 

We defend an evidentiary account of internalism that entails both mentalist and access criteria for epistemic justification. The evidentialist thesis holds that what makes one justified of holding a particular belief is the evidence that one possesses \cite{mittag2011evidentialism}. We hold that agents have internal access to one’s evidence through mental states. As such, justificatory factors are not merely mental states but mental states to which agents have exclusive, internal access. It is the joint endorsement of the mentality and the access theses that define our brand of internalism. We call the explication of evidentialism through accessibility and internalist criteria reflective knowledge. Agents in possession of reflective knowledge have access to the rational conditions or bases that justify their beliefs. This does not imply that justified beliefs are infallible or that the justificatory status of an agent’s network of beliefs is complete. It implies, however, that agents that possess reflective knowledge can evaluate knowledge claims and beliefs relative to different standards of knowledge. 

In conformity with the foregoing, we define internalism as follows: \\

\textbf{Internalist Justifiedness (IJ)}: \textit{A belief can amount to knowledge only through backing of reasons adduced as premises}. \\

As a brand of internalism, we define reflective knowledge as follows: \\

\textbf{Reflective Knowledge (RK)}: \textit{``The reflective knowledge agent not only holds a justified, true belief but also possesses an awareness of the reliability of their belief-forming process, allowing them to recognize and affirm that they know."}\cite{sosa2011reflective}
\\

Let’s flesh out these affirmations by way of an example. One cannot justify the existence of transfinite numbers unless one understands the proof. The proof constitutes the series of logical steps from which the conclusion follows. A rational agent may choose to unreflectively accept the conclusion by deferring to the reliability of the proof or she may choose to understand the rational steps the proof takes. Since the proof for transfinite numbers rests on axioms, other rational agents may choose to deny the axioms. Whether or not we accept the axioms, the following hypothetical remains ironclad: if we take the axioms to be true, the proof guarantees the conclusion of the existence of transfinite numbers. This is the standard of deductive inference. To take a more familiar example, the Pythagorean Theorem, whose applications abound in physics, engineering, cartography, construction, and agriculture, has a wide number of proofs. The truth of the theorem rests on the deductive proof. Independent human reasoners may verify the correctness of the proof, but it is its veracity that enables its reliable application across disciplines and practices. While the rational basis for the proof is independent of human reasoners, it is because human reasoners can independently evaluate the reasoning as correct that we accept it to be true. In virtue of what do human reasoners evaluate the proof? We answer: in virtue of reflecting on and understanding the logical steps. While philosophical analysis concerning the definition of reflection and understanding is ongoing, we endorse the internal access thesis expounded on above. 

The deductive examples we have presented extend also to empirical knowledge. It is because independent human reasoners evaluate the predictive power of General Relativity against observation that the prevailing scientific consensus accepts the theory as broadly corroborated. In short, the reflective standard reigns as the conferrer of truth or veracity. This reflective standard imposes internal consistency constraints as well as empirical correspondence constraints. Our knowledge must not possess logical contradictions and must rest on deductive valid reasoning and inductively strong empirical bases. Anything short of the reflective standard reduces to heuristic or deferral to reliable processes. As we will argue later, rational agents must also reflect on the reliability of knowledge they accept without walking through the onerous reasoning steps. While rational agents can afford to adopt a doxastically agnostic attitude to many knowledge claims, they cannot afford to do so with respect to all, especially those knowledge claims upon which interlocking social processes depend. 

Reflective standards of justification as we have characterized them entail normativity by imposing upon human agents context-sensitive epistemic duties. In their professional roles, to take one epistemic context among many, human agents have an epistemic duty to apply the highest rational standards. As citizens, they have the epistemic duty to stay informed and evaluate their sources of information. Human agents are therefore tasked with applying reason within contextual demands of the social processes in which they participate and on which the collective depends. As we will argue, this does not mean that every individual must reason rationally through every belief they hold. Necessarily, human agents outsource rational standards to the broader collective. But it is because rational standards are rigorously applied by individuals in circumscribed domains that human collectives produce broadly reliable bodies of knowledge. It is because a proof for the Pythagorean Theorem was discovered that generations later we can reliably apply it to engineering. 

The affirmation of the internalist standard of justification is not a trivial thesis. We advance it as an explanatory premise to the phenomenon of the accumulation of human knowledge \cite{odling2011ecological}. It is because human collectives reflect on the grounds on which their knowledge rests that they modify and improve upon it unlike proximate species and presently, state-of-the-art AI. By propounding the internalist epistemic standard as the highest standard of knowledge we do not concomitantly heave humans on a pedestal above other intelligences. Rather, we hold the standard of reflective knowledge as true, independent of whether humans or other intelligences possess it or not. It is, consequently, the hallmark of general intelligence insofar as it entails a standard we can generally apply to an unlimited array of problems in order to guarantee or approximate truth rather than merely outsource it to some reliable process. 

\subsection{Externalism}

The internalist conception of justification we have defended faces competition from an entirely different standard of justification termed externalism. Contrary to internalism, externalist theories of justification place the onus of justification in the sources of knowledge rather than access to the basis of justification. As such, externalist theories deny either or both the sufficiency and necessity conditions of internalist justification. The leading externalist theory, developed and championed by Alvin Goldman, is termed reliabilism. Reliabilism holds that a subject’s beliefs are justified if they are caused by reliable belief-causing processes \cite{goldman2020reliabilism}. Reliabilism differs from internalism in that it absolves the agent from having reflective access to the bases that render their beliefs true or likely to be true. The justifiedness of beliefs instead depends on the causal history through which subjects acquire their beliefs \cite{goldman1979justified}. Goldman shifts the emphasis from internal states where justification amounts to a set of reasons or arguments on which agents ground their beliefs, to processes, properties or causes that confer justification on beliefs independently of the agent’s understanding of those bases. 

Goldman explicates reliabilist criteria as follows: 

\begin{quotation}
“Which species of belief-forming (or belief-sustaining) processes are intuitively justification-conferring? They include standard perceptual processes, remembering, good reasoning and introspection. What these processes seem to have in common is reliability: the beliefs they produce are generally true. My positive proposal, then, is this. The justification status of a belief is a function of the reliability of the process or processes that cause it, where (as a first approximation) reliability consists in the tendency of a process to produce beliefs that are true rather than false” \cite{goldman1979justified}
\end{quotation}

We accept Goldman’s insight that reliability is a necessary condition for knowledge and a fortiori collectively distributed knowledge. In fact, we contend that reliabilist conditions make reflective knowledge possible. It is because evolved cognitive processes such as perception, attention, memory, and non-linguistic inference or mental modeling reliably track the truth, that reflective knowledge becomes possible at all. In other words, reliable cognitive competencies and information processing form the condition of possibility for knowledge. The view that our knowledge is broadly non-reflective is known as radical externalism \cite{morton2008guide}. However, while accepting the necessity of reliability as a condition for knowledge, we hold that reliable information processing systems are insufficient for reflective knowledge. Reflective knowledge adduces upon merely reliabilist knowledge additional standards such as correct rules of reasoning through explicit arguments. Because humans codify the world symbolically and reflectively refine those symbolic representations in light of additional evidence, human knowledge remains an open and ongoing project. As such, reflective knowledge explains the accumulation of human knowledge in a way that mere reliabilism cannot. Along these lines, we adopt Ernest Sosa’s distinction between animal and reflective knowledge, where animal knowledge broadly denotes reliable belief-causing processes, while reflective denotes knowledge that additionally adduces justifications for the methods of reasoning themselves \cite{sosa2007virtue}. Accordingly, we maintain a logical distinction between causation and justification and do not reduce the latter to the former. 

Having affirmed the necessity of reliability as a condition for knowledge, we extrapolate Goldman’s reliabilist theory to social processes. Goldman neglects to mention institutional processes as candidates of reliable belief-causing processes, presumably because he is primarily concerned with individual epistemic agents. However, adopting Goldman’s general view that justification is partly a function of reliability, we argue that the greater part of knowledge, even in the case of individual agents, is caused by institutional processes. For example, students derive most of their abstract, descriptive knowledge from textbooks, classrooms and broadly educational institutions. While that knowledge has been accumulated through a historical application of rigorous logical and empirical standards by communities of scientists, students inherit that knowledge by trusting the chain of cultural and social transmission. This observation persuades us to distinguish between two broad categories of reliabilist sources of knowledge: a) evolved cognitive competencies on the one hand and b) cultural processes of knowledge production and transmission on the other. Both of these sources of reliabilist knowledge entail fallibility. That is to say, neither guarantees the transmission of truth. The recognition of fallibility induces the incremental imposition of additional constraints on knowledge that allows the body of knowledge to grow rather than remain static. Because cultural processes of knowledge transmission for the most part filter out false theories, the preponderance of false theories over approximately true theories is not apparent \cite{psillos1996scientific}. Consequently, we argue that the locus of knowledge growth resides with internalist standards. Internalist justification, we argue, necessitates the refinement of knowledge standards that lead to more robust, stable, and growing bodies of knowledge. While evolved cognitive competencies constitute a paradigmatic set of reliable belief-causing processes, we will focus on institutional processes as imposing additional constraints on knowledge through division of rational labour and application of domain-specific standards on different areas of knowledge. We term these latter processes institutional epistemology. In developing an institutional epistemology, we incorporate reliability as a necessary standard of transmission, encapsulated by Ernest Sosa's principle of the criterion: 

\textbf{Principle of the Criterion (PC):} \textit{``Knowledge is enhanced through justified trust in the reliability of its sources" } \cite{sosa2007virtue}. 

\subsection{Institutional Epistemology}

Having propounded the internalist standard yet argued that the externalist standard is indispensable due to the limited rational resources of individual agents, we advance a normative conception of epistemology operant at the level of social collectives that necessarily combines internalist and externalist standards. In order for knowledge as a collective project to both scale and grow, both standards are required to be applied in accordance with prescriptive criteria. 

Institutional epistemology concerns knowledge as a collective project produced through processes of social cooperation rather than a single individual’s relationship to their beliefs. Collective rationality refers to the distribution of knowledge across communities of practice. In advancing a theory of institutional epistemology, we recognize that, in many respects, the traditional conception of epistemology as oriented toward maximizing the rationality of an agent’s beliefs constitutes an impossible standard for knowledge. This insight is motivated by Goldman’s conception of reliabilist forms of justification. However, contrary to Goldman, we pursue reliability as a function of collective or social epistemology. Individuals pursue and produce knowledge as part of cooperative social schemes that rely on a stratified division of rational labour. Collective rationality holds that the accumulation of knowledge occurs through social communities of practice. 

Two definitions of collective rationality pervade the literature: collective decision and rationality at the scale of collective entities. A collective decision for example may be arrived at through communication and deliberation within a group. Moshman and Geil \cite{moshman1998collaborative} provide evidence that collaborative reasoning in groups far outperforms individuals on the selection task, a paradigm of hypothesis testing. Collective entities on the other hand include institutions, organizations, and states. Decision processes and controls on information-flows within such entities are delegated and distributed. A corporation, for example, can be considered to behave rationality when its decisions subserve its profit-maximization bottom-line. A state can be considered rational if its actions and decisions subserve the collective ends of its citizens \cite{Kirman2010}. 

We subscribe, therefore, to a theory of bounded rationality with respect to the epistemic horizons of individual actors predicated on cognitive and temporal limitations. A corollary of this view is that individual persons must outsource a great deal of their beliefs to reliable belief causing and transmitting mechanisms. It additionally implies that epistemological standards are instantiated at the level of collective action and span across organized collectives such that individual agents acquire part of their epistemic warrant from collective standards. 

However, we also recognize that collective standards can atrophy for various reasons such as malpractice, poor ethics and misallocation of incentives, which in turn requires individual actors to play a role in evaluating social mechanisms of belief transmission. On this view, the epistemic duties of individuals are twofold: a) they must play a role in the division of rational labour by applying internalist standards to the subset of knowledge tasks to which they contribute and b) they must continually evaluate the mechanisms whereby they source knowledge evaluated on reliabilist criteria. Epistemic duties imply that normative epistemic standards are necessary for the epistemic health of collective rationality and institutional knowledge. When individuals apply the highest epistemic standards to their own domain while evaluating the degree to which these epistemic standards have been upheld in other domains, they increase the likelihood of maximizing truth and minimizing error across chains of epistemic dependence and knowledge transmission. We theorize that this reciprocal influence between collective and individual rationality constitutes a necessary normative dynamic of reciprocal constraint between local-to-global and global-to-local justificatory directionality on knowledge production that robustly staves off the threats of epistemic vices and fallibilism. 

We argue that individual and collective epistemological standards are therefore deeply intertwined. We theorize two necessary conditions for their mutual dependence: a) individuals must apply robust rational standards to the knowledge domains where they exert professional authority and b) individuals must apply ongoing evaluative meta-standards to the wider mechanisms of knowledge production and transmission. The latter condition entails minimally that rational agents evaluate their sources, but stringier constraints would require that they consider a host of interconnected factors that holistically confer reliability on the requisite knowledge. This view of epistemic justification is outwardly normative in that it imposes epistemic duties on individuals as social actors that participate and benefit from social processes and collective action. 

We supplement our internalist account of knowledge with a thesis of collective justification. Collective justification entails that knowledge distributed across agents and institutions can only be sufficiently justified at the level of collectives. To return to the example we provided earlier, one cannot justify the existence of transfinite numbers unless one understands the proof. Understanding the proof constitutes the highest standard of justification with respect to a mathematical theorem. Weighing all the relevant evidence with respect to an empirical proposition constitutes the highest standard for justifying that proposition. Bounded rational agents with limited cognitive resources can only justify a fraction of their knowledge by meeting full-fledged rational and reflective standards. As such, bounded rationality necessarily entails that rational agents outsource the application of rationality to other agents. In doing so, they forfeit the application of internal, reflective standards to the vast majority of their knowledge by broadly trusting that other agents have applied those standards in their area of expertise. In the ideal justificatory scenario, the division of rational labour entails that social rational agents do their part in increasing the net justificatory status of collective knowledge. Collective knowledge consequently imposes deontic constraints on rational agents to apply the highest rational and evidentiary standards in their area of expertise. It is only through the satisfaction of individual deontic epistemic duties that individual agents can trust the transmission of knowledge in areas where others bear the duty of epistemic justification. 

The justification of knowledge at the collective level, therefore, involves a distributed dynamic of internal standards and reliable transmission mechanisms such that total knowledge is justified at the level of collectives rather than mere individuals. This necessitates a new standard for collective justification that balances the constraints of individual rationality with collective rationality. In order to produce a scheme where communities of practice and groups of communities of practice reliability transmit knowledge, bounded rational agents must fulfill their epistemic duties. 

In order to ensure the logical consistency of these two theses, we hold that all internalist justification entails a degree of externalist outsourcing of aspects of knowledge. This means that reliabilist and internalist criteria of justification exert mutual constraint on the constitution of broadly justified bodies of knowledge. Structural engineers must rely on the correctness of equations in order to perform their job, while everyday commuters must rely on the instantiation of structural engineering standards when crossing the bridge. The reliability of infrastructure, therefore, relies on accumulated social standards of expertise. Chains of accountability in the application of standards are distributed within organizations and across legal bodies that enforce and oversee their adherence. How have standards come to be? Individual structural engineers may learn the standards unreflectively as a set of professional rules to which they must adhere. At the same time, the body of standards evolved through a social process of trial and error and the testing of reasons against practice. The structural engineer can reflect on the reasons that justify the standards and adduce reasons for modifying the standards in light of changing practices, technology, and new evidence. Standards of truth and accuracy within special domains, therefore, are open to revision through reflective evaluation of the reasons upon which they stand and the evidence available to collections of human reasoners. 

In light of the above arguments, we propose a new thesis that posits the mutual dependency of internalist and externalist standards of justification. We articulate our thesis of collective justification as the conjunction of three individually necessary theses that jointly meet necessity and sufficiency criteria for collective justification: 

\textbf{Internalist Standard (IS)}: Reflective knowledge is necessary to establish truth within a domain of study. 

\textbf{Externalist Standard (ES)}: Reliable belief-causing and belief-transmitting processes are necessary for inter-domain knowledge to grow and support complex collective action. 

\textbf{Normative Standard (NS)}: Because both reflective and reliabilist standards are fallible, subject to error, bad reasoning, and revision in light of new evidence, it is necessary that rational agents cultivate epistemic virtues viz a vis. IS \& ES that increase the likelihood of maximizing the discovery and transmission of truth. 

In affirming the joint truth of these three propositions, we deny the following claim held by some philosophers: 

\textbf{Disjunction Clause (DS)}: Any theory of knowledge must be internalist or externalist (exclusive or). 

\subsection{Epistemology \& LLMs}

In light of the remarkable learning capacities of LLMs, the question naturally arises whether they have knowledge of the kind we have been discussing. 

The rise of connectionist artificial intelligence marks a qualitative shift within cybernetics. Traditional computers, GOFAI, and the internet can be construed as technologies that under the right conditions enhance collective rationality by increasing access to information and expand possibilities for cooperation. At the same time, they equally present risks to collective rationality by enabling communities of practice that diverge from and subvert rational standards. For example, increasing online social interaction can dislodge individuals from broader social norms through self-reinforcing communities that operate untethered from tried and tested social standards. This is attested by the phenomenon of echo-chambers . Both the benefits and risks within traditional computers, networks and GOFAI reside with structural changes to collective intelligence. Computers increase the computational power of individuals, the internet increases the possibilities for information diffusion and geographically untethered communication, and GOFAI automates reasoning processes. 

Connectionism, on the other hand, signals the emergence of algorithms that learn domain-specific skills on large bodies of data. The emergence of Generative AI and LLMs marks an inflection point in the evolution of artificial intelligence in virtue of replicating the abilities of brains to generalize from available information. 

Here, we advance a formal distinction between the processes through which an ANN learns and the processes through which a human learns and reasons. We believe that articulating these differences is not trivial in light of the remarkable capacities of current LLMs to learn from data and surpass human capacities in speed of linguistic generation on the one hand, and a range of cognitive tasks on the other \cite{luo2024large, mcclure2024ai, pymnts2024anthropic, zhang2023circumstances}. The underlying architecture of LLMs structurally reproduces the general contours of bottom-up learning and top-down inference instantiated by human brains. Broadly speaking, ANNs learn by processing large amounts of data through a hierarchical network structure consisting of layers of artificial neurons \cite{smolensky1987connectionist, dayan1995helmholtz, sun2001implicit, lecun2015deep}. In supervised learning, the model error is computed by comparing the network output to ground truth or expected values \cite{bengio2016deep, ghosh2017application, janochaczarnecki2017loss}. In self-supervised learning, the model constructs supervision signals from the data without the need for human labels to optimize predictive tasks \cite{lecun2021selfsupervised}. In reinforcement learning, where learning occurs through reward signals from the dynamical interaction between an agent and its environment, the target state is defined by a policy of cumulative reward \cite{sutton2018reinforcement}. Whether the target derives from human labels, the internal structure of the data, or a reward policy, all deep learning minimizes a loss or error function defined over some target signal as the core mechanism of learning. The computation of error, whose function varies across tasks, enables the retroactive adjustment of connection weights through backpropagation \cite{lecun2015deep, rumelhart1986learning}. The iteration of this process of forward passes and backpropagation minimizes the error and increases the accuracy of the model. When scaled to billions and trillions of parameters in conjunction with sophisticated architectures such as Transformers, whose self-attention mechanism processes textual data strings in a parallel fashion by converting the input into query-key-value vectors \cite{cheng2016long, vaswani2017attention}, LLMs display remarkable pattern recognition and generalization capabilities. In other words, through the explicated learning process, the model acquires an internal representation of the data distributed across the neuronal connection weights \cite{mikolov2013distributed, armenta2020representation}. The nature of these representations is a matter of debate, though the parallel distributed processing thesis in cognitive science and the philosophy of mind analogizes artificial neural networks’ distributed representations with those of biological neural networks in biological brains \cite{rumelhart1986parallel, patterson1989connections, mcclelland2003parallel}. In this regard, AI qualitatively differs from symbolic AI in that it holds internal representations of the data rather than representations manually coded by humans. The model, therefore, learns of its own accord and the role of the human designer does not lie with manually coding rules but adjusting the model hyperparameters such as the number of neurons, number of hidden layers, loss formula, optimization algorithm, learning rate, and so on, to optimize performance \cite{zou2008overview, lecun2015deep}. The model does not have the capacity to design itself, though self-regulating ANNs and LLMs that can adjust their own hyperparameters have been proposed \cite{wang2024llm}. As we have described them, ANNs possess a kind of knowledge that structurally resembles what we have termed animal knowledge, following Sosa \cite{sosa2011reflective}. ANNs can generalize through a form of bottom-up processing structurally inspired by the hierarchical network architecture of brains where individual neurons process information at scale through, roughly speaking, binary activation decisions \cite{rumelhart1986learning}.

However, ANNs and LLMs lack the following necessary dimensions of the kind of robust knowledge we have been touting: 

\begin{enumerate}
    \item Access to ground truth
    \item Reflective knowledge and reasoning capacities 
\end{enumerate}

Despite proposing that LLMs possess or instantiate a kind of animal knowledge, we recognize that LLMs contain robust representations of the data upon which they have been trained, whatever the character of those representations may be (coarse-grained symbols or merely distributed patterns). However, LLMs do not have access to the basis of their knowledge, namely to the reasons as to why their knowledge is well-founded. In other words, they constitute reliable transmission mechanisms of already well-reasoned information \cite{felin2024theory}. Further, they have no access to ground truth in two distinct ways: a) they rely on samples controlled by humans and b) they are incapable of evaluating evidence as a form of ground truth \cite{lake2017building}. The potential usage of ANNs and LLMs specifically for scientific discovery \cite{si2024can} still requires the explainability of the patterns by human reasoners. These shortcomings severely undercut the epistemic prowess of ANNs and LLMs. This is evidenced by the persistence of hallucinations. Broadly speaking, generative AI and LLMs as their proper subset suffer from the hallucination problem by exhibiting a tendency to sometimes generate seemingly plausible outputs that are in fact nonsensical or nonfactual \cite{huang2024survey, rawte2023troubling}. In LLMs, hallucinations stem from a variety of causes including pre-training and alignment data, the training process itself such as next-token prediction, and lastly from the decoding or inference process where the model generates text sequences based on its internal probability distribution \cite{huang2024survey}. Further, some researchers argue that the hallucination problem is a structural feature of generative models that arises from computational limits and the inherent incompleteness of the prior (i.e., the training data) and therefore cannot be entirely eliminated \cite{banerjee2024llms, xu2024hallucination}. If hallucinations are structural as some argue, it imposes limits on their precision and reliability as truth-transmission mechanisms.

Despite the generalization and pattern recognition prowess of current ANNs, ANNs do not know why they know as humans do. They may reliably generalize from data, but cannot justify their knowledge with respect to the reflective standards we have explicated. To possess this evaluate capacity, ANNs need to be able to represent the conditions through which they learn to themselves. That is to say, they would require internal reflective access to their representations. This capacity is a feature of human minds that at present escapes engineering because it eludes mathematical formalization. It is because humans can justify the very reasoning processes upon which they scaffold knowledge that they come to know the truth about a phenomenon or mathematical theorem rather than merely relying on a set of reliable belief-causing processes. Finally, because LLMs are at the mercy of the sampling data of reflectively evaluated human knowledge, they fundamentally reduce to reliable transmission mechanisms with pattern-recognition capacity. Put differently, the epistemic performance of LLMs is as good as the epistemic status of the data on which they have been trained. If LLMs are fed inaccurate data and poorly reasoned bodies of knowledge that do not track the truth, they will unreflectively reproduce those errors.

\subsection{LLMs \& Reflective Knowledge}

Arguing that LLMs do not have epistemic access to their knowledge base carries important and far-reaching implications for collective rationality and for public and educational institutions, as well as information-oriented organizations. As we will highlight, LLMs stand to reshape the landscape of information transmission and retrieval currently reliant on databases, search engines, and non-deep learning-based retrieval and recommendation algorithms. We will outline the role of these institutions and organizations as stewards of epistemic virtues, capable of guiding collective norms toward minimizing epistemic threats and maximizing epistemic virtues.

Because LLMs lack reflective knowledge, they cannot as yet adduce epistemic justifiedness to bodies of human knowledge but merely reliably transmit already rationally justified bodies of knowledge. While LLMs cannot produce reflective knowledge, namely knowledge with reflective justification, it does not mean that LLMs cannot be utilized in workflows of knowledge production and novel pattern recognition and hypothesis generation. It does, however, mean that use of LLMs for knowledge production requires human reflective explainability for those new hypotheses to count as knowledge. While LLMs and ANNs more broadly can detect novel patterns, determining whether those patterns constitute cogent hypotheses rather than noise and gibberish rests on the human reasoner. That is to say, human reflective validation remains currently necessary to buttress novel AI predictions with explainable ancillary hypotheses \cite{ludwig2024machine}. This implies that human users need to be aware of the threats of outsourcing reasoning tasks to LLMs at the expense of foregoing their epistemic duties to apply reasoning and the highest rational standard to relevant tasks. It also conversely implies that LLMs carry opportunities on top of epistemic threats to increase human understanding by revolutionizing information transmission, seeking, and retrieval.

Advancing the view that LLMs do not have reflective knowledge implies taking a position about their information transmission and production role in human–LLM interaction tasks, on the one hand, and organizational and institutional processes on the other. We hold that reliance on LLMs for knowledge transmission and decision-making requires model transparency as well as the oversight of human evaluators and decision-makers. In light of the foregoing arguments about the epistemology of LLMs, we will highlight a host of epistemic threats posed by LLMs for both individual rational actors and collective rationality as a social project.

\subsection{XAI and Hybrid Human-AI Reasoning}
In this section we consider counterarguments to our account of knowledge as balancing reflective and reliabilist standards stemming from pragmatist conceptions of knowledge and claims that LLMs produce proto-reflective knowledge. 

While pragmatist definitions of knowledge are numerous, we will take Dewey's definition as warranted assertability as a common example \textcite{dewey1941warranted}. We should note that our conception of knowledge shares important attributes with pragmatism: we embrace the view that reflective knowledge is fallible, while making no pretense at arriving at certainty. We reserve \textit{certainty} exclusively for valid deductive inference, which is agnostic about the truth-value of the premises. Consequently, we hold that empirical beliefs are inherently probabilistic and justify degrees of credence in rational agents analogous to Bayesian inference. 

It is worth considering whether self-supervised learning models (SSL) like Bidirectional Encoder Representations from Transformers (BERT) instantiate proto-reflective reasoning. Autoencoders like BERT involve encoder-decoder architectures that excel at natural language understanding (NLU) \cite{devlin2018bert}. Consequently, it is crucial to explicitly state why SSL models do not possess reflective knowledge \cite{gui2024selfsupervised, liu2021ssl}. Even though an SSL model learns without human-annotated labels, it produces knowledge of the type we have identified as animal knowledge. While an autoencoder like BERT can outperform humans at linguistic comprehension in terms of speed, volume, and statistical regularities, it is beholden to the data on which it is trained \cite{simmons2023garbage}.  Consequently, it learns textual statistical regularities without knowledge of real-world, causal relationships that humans possess \cite{felin2024theory}. Causal AI, a growing subfield in AI that aims to model ANN causal inference and discovery through directed acyclic graphs (DAGs) and structural causal models (SCM) requires access to a known causal structure or predefined assumptions \cite{pearl2009causality, scholkopf2021causal}. Despite outperforming biological brains at specialized tasks, BERT and other state-of-the-art (SOTA) SSL models resemble animal knowledge in virtue of the fact that their knowledge is exhaustively explained by recourse to externalist processes of pattern recognition. Reflective knowledge requires, in addition to automatic generalization capacity and additional heuristics, that an agent internally represent the bases in virtue of which their generalized patterns count as knowledge. This necessitates that reflective agents possess a global view of the state of their reflective knowledge that bestows internal access to the evaluative standards that confer warrant on their beliefs. To give an example, a human agent can possess internal access to how a differential equation is algebraically justified but no internal access to how their perceptual system renders certain colors. However, they can in principle gain reflective knowledge of the latter processes through empirical investigation, in which event, they come to internally represent the causal sequences that produce observed perceptual effects. In other words, reflective agents maintain internal representations not only of first-order knowledge, but also represent to themselves the conditions under which such knowledve counts as knowledge. It is on the latter basis that reflective agents revise the justification of their beliefs. Explicating the conditions that make reflective knowledge possible requires delving into theories of mental contents, second-order awareness, and higher-order thought, which lie beyond the scope of this paper. 

The field of XAI aims to make deep learning based outputs explainable through post-hoc methods. XAI's theoretical aspirations of making black-box machine learning explainable would help justify the integration of black-box models within high-stakes decision making \cite{lipton2018mythos, murdoch2019interpretable, ruambo2019towards, minh2022explainable}. Remarkable rates of predictive success without understanding how the model reaches its decisions pose a host of ethical problems for integrating AI models within socially consequential decision-making \cite{rudin2019stop}. Explainability, therefore, is touted as a variable that aims to convert the epistemic opacity of deep learning models into epistemic transparency \cite{minh2022explainable}. Jointly, predictive success and epistemic transparency make for the responsible deployment of AI in institutional and organizational processes. However, XAI remains highly aspirational due to the limitations inherent with post-hoc explanations. Post-hoc models such as LIME \cite{ribeiro2016why} and SHAP \cite{lundberg2017unified} that aim for plausible explanations suffer from the underdetermination problem. The \textit{underdetermination of theories by data} thesis holds that multiple, equally plausible explanations can account for the prediction \cite{newton-smith2017underdetermination}. The true test of explainability would be if post-hoc XAI could predict the future behavior of a model. Since this is not yet the case, XAI has a long way to go to make its theoretical aspirations a reality. However, if we hypothetically assume the success of XAI, it is worth considering whether it would alleviate the epistemic worries that we have raised. Even if, in theory, we could explain the processes whereby a deep learning model reaches its decisions or gain insight into its coarse-grained representations, this would not confer reflective justification upon deep learning models and the same problems of human epistemic erosion would persist. However, it would open up such models to evaluation through human reflective capacities, and XAI-assisted performance would improve the overall transparency of decisions. These considerations are currently theoretical and the post-hoc nature of XAI ensures that the black-box problem is intrinsic to deep learning models and not exhaustively eliminable \cite{bordt2022posthoc}. 

Hybrid Human-AI intelligence (HHAI) proposes that humans and AI systems can collaboratively solve complex tasks by complementing each other’s strengths, combining human judgment, contextual understanding, and ethical reasoning with the scalability, speed, and pattern recognition capabilities of AI \cite{dellermann2021hybrid, hao2025humancentered, jarrahi2022hybrid}. HHAI encompasses XAI since XAI-assisted performance can be shown to yield improved performance \cite{senoner2024explainable}. Dellermann et al \cite{dellermann2021hybrid} find that HHAI systems outperform humans or AI alone in task performance, adaptability, learning efficiency, and decision quality, particularly in complex, uncertain, or dynamic environments. They advocate for the design of socio-technical systems in which humans and AI agents co-evolve, collaborate continuously, and share decision-making responsibilities to achieve superior outcomes neither could attain alone \cite{dellermann2021hybrid}. We agree that HHAI is an inevitable outcome of human-AI interaction (HAX) just as HCI was an inevitable outcome of the advent of mainframe and later personal computers. As in HCI so in HAX the interaction outcomes involve a redistribution of agency. Furthermore, our thesis is entirely consistent with HHAI broadly conceived, but agnostic about the evolutionary shape of HHAI. The assumption underlying many HHAI proponents is that a hybrid intelligence produces an additive outcome without consequences for the human agent. Conversely, we hold that hybrid intelligence without the requisite epistemic and interaction norms can result in the atrophying of human reflective knowledge. However, properly steered HAX also holds the potential for conferring epistemic benefits as we will recount in the section "maximizing epistemic virtues".

\subsection{Epistemological Threats}

Human-LLM interaction presents both epistemological threats and opportunities. Threats distribute across four categories: erosion of epistemic norms, erosion of incentives to learn, and diffusion of ignorance and transmission of errors. The first two concern the epistemological practices of individual reasoners, whereas the latter two concern the scaled effects of the former on collective rationality. 

While we have argued that reflective knowledge forms the highest standard of knowledge, we had additionally asserted that reflective knowledge depends on complex social scaffolding as antecedent conditions conducive to its instantiation. Consequently, we define reasoning as a social process contingent upon the following social conditions: \\

\begin{enumerate}
\item Reasoning occurs within socially-sanctioned social contexts 
\item Reason responds to incentives within a social structure
\item Reasoning rests on a dialogical basis of mutual understanding between human reasoners (Habermas 1992)
\item Reasoning constitutes a type of social action 
\end{enumerate}

We outline the following epistemic threats posed by LLMs to individual human reasoners:

\textbf{Erosion of epistemic norms:} Epistemic norms refer to the practices through which agents evaluate the sources and status of their beliefs as well as their interconnectedness with respect to available evidence. In the ideal epistemic scenario, reasoning agents seek conceptual understanding and corroboration of their beliefs against available evidence. Collective epistemology imposes duties on rational agents to procure the best available evidence with respect to some socially consequential task for which they are responsible. It furthermore imposes constraints on epistemic practices from basing beliefs and inferences on inconclusive or inadequate evidence and/or incomplete conceptual understanding. Reliance on LLMs on these tasks can lead to the erosion of these norms of evidence gathering and conceptual robustness. 

\textbf{Erosion of learning incentives: }Learning refers to the process of obtaining knowledge of some domain that aims toward the highest level of comprehension and practical mastery. Incentives and motivations for learning are intimately tethered to evaluation outcomes on the short term and social outcomes such as entering a profession in the long term. When learning agents outsource their comprehension and application of a knowledge domain to an LLM, they forego the process of acquiring personal mastery. This, in turn, increases reliance on LLMs for comprehension and task completion. This reliance additionally can erode dispositions and habits of learning practices given the long-term reliability of LLMs. 
 
\textbf{Diffusion of ignorance:} The erosion of epistemic virtues and incentives to learn can increase the net ignorance of a population. When people know less about domains of knowledge, whether it be their profession or general knowledge, the process of refining knowledge also proportionally diminishes. Therefore, the erosion to seek and justify individual knowledge, can lead to the spread of ignorance and consequently lower performance results by human agents on tasks that require specific domain expertise as well as practical wherewithal. 
 
\textbf{The transmission of error:} Related yet distinct from the diffusion of ignorance concerns the transmission of error across systems, organization, and social networks. The transmission of error already has proven to be a problem within social media and online communities where false information propagates rapidly across users. More broadly, the issue here concerns the transmission of error in professional and high-stakes settings. Reliance on LLMs for factual as well as well-reasoned information could lead to human reasoners and decision-makers incorporating error in their tasks and decision-making. While error can be transmitted reliably through machine processing, the erosion of epistemic virtues and comprehension in human reasoners can in turn erode their ability to conduct oversight on error-prone autonomous systems. 

In conjunction, these epistemic threats are causally tied to the social scaffolding that fosters the practice of reason across populations. If reliance on LLM-interaction becomes the norm, the conditions that foster reasoning through dialogical engagement with other human reasoners, the process of gaining individual comprehension, and the reward systems that sanctions these practices will also erode. 

In line with these arguments, we will outline the consequences of these threats for collective rationality posed by human-LLM interaction. We distinguish between two types of epistemic consequences: 1) local epistemic threats and b) global epistemic threats. 

\textbf{Local Epistemic Threats (LETs)}: The outsourcing of learning, understanding, and reasoning by individual human agents to LLMs resulting in the transfer of reflective epistemic practices to reliable LLM outputs. 

\textbf{Global Epistemic Threats (GETs)}: The scaled impacts of outsourcing collective reflective knowledge and reasoning processes to LLMs resulting in the erosion of collective justifiedness. 

These threats entail reciprocal causal interactions between the local application of reason by the human agent and global configurations of collective rationality are reliant on the division of rational labour. We term these local-to-global and global-to-local effects on epistemic norms. 

\textbf{Local-to-Global (LTGs)}: Causal diffusion from local epistemic threats to collective justification whereby a critical mass of lower epistemic standard adoption reduces the net justifiedness of human bodies of knowledge. 

\textbf{Global-to-Local (GTLs):} Top-down causal effects on individual human agents from the collective adoption of new epistemic norms that lower the bar of epistemic justifiedness.  

We have argued that LLMs outsource their knowledge to trustworthy information that humans have already reasoned through and thereby sample from the pool of collective human rationality. To the extent that LLMs transmit knowledge that has been epistemically verified through correct reasoning procedures by humans, it remains a reliable transmitter of information. However, outsourcing problem-solving to AI at scale implies the longitudinal erosion of the justificatory status of the pool of human knowledge. If enough human agents and by extension institutional processes offset their problem solving and reasoning processes to AI, then AI will have less reliable information from which to pool its answers. Consequently, If access to LLMs erodes incentive for humans to learn, expand their understanding, and reason by first principles, then the scaled outcome could lead to large-scale outsourcing of reasoning processes and decision making to autonomous artificial agents whose bases of inference is beset by the black-box problem of AI interpretability. 

Human reason rests on a delicate and incrementally scaffolded process of socialization whereby the developing child is initiated into literacy and a condensed period of learning through which it acquires culturally accumulated knowledge. It is because discoveries and inventions of the past are transmitted to succeeding generations that the latter can incrementally contribute to the growing inheritance of human knowledge.The advent of digital electronic computers during the middle of the twentieth century, marked a fantastic cultural shift that accelerated cultural accumulation. Steadily improved performance of the digital computer through integrated circuitry as per Moore’s law has exponentially augmented the capacity for information generation, storage, retrieval, and processing, in effect shifting the role of the human agent toward higher intellectual tasks. This remarkable computational capacity expands human reasoning by exponentially increasing the computational range, which in prior epochs was confined to human manual calculation and non-electronic mechanistic extensions. Exponentially greater computational capacity enables the automation of serializable tasks on the one hand, and scaled mathematical modelling to previously unattainable levels on the other. 

However, Generative AI represents an inflection point in this trajectory of the augmentation of scientific enterprise and cultural accumulation in virtue of the fact that it threatens to automate or erode reflective reason. Since we have argued that current AI does not possess reflective reason, we have emphatically maintained that reflective reason cannot be as yet automated. However, the reliability of LLMs and Generative AI more broadly produces large scale scenarios where the incentives on which reflective reason rests will erode. In the foregoing we will present solutions to these problems as well as ways that human-LLM interaction can be properly framed to produce benefits rather than costs to reflective reason. 

Already, LLM-powered chatbots are seeing large-scale adoption by users and adoption by institutions \cite{dekeyser1996exploring, sood2024exploring}. LLMs are being adopted in education \cite{li2023adapting, ng2024llms}, medicine \cite{sayin2024can, thirunavukarasu2023llm, wang2024llms}, law \cite{almeida2024exploring, lai2024large, cheong2024ai}, finance, to name a few yet consequential domains of adoption. Just as the adoption of computers and ICTs, the adoption of LLMs and other Generative-AI technologies are likely to reshape the landscape of labour, human agency, and organizational information management practices. The effects on individual adoption are likely to remain more amorphous, though use of chatbots as personal assistants is on the rise. These adoption trends in information seeking, management, and retrieval will likely follow Zipf’s law of least resistance. These trends fit the view of LLMs and chatbots as enhancing human plans institutionally and individually. 

In outlining these threats, it is important to consider limitations and alternative evolutionary scenarios. Assuming expediency of reliable information procurement overrides full-fledged reasoning and reflection, it means that if the pool of reliable knowledge stagnates due to mass outsourcing of reasoning tasks to LLMs, it will thereby create new incentives to practice full-fledged reasoning in relevant contexts. The looping of LLMs into personal workflows and institutional processes may cause system adjustments that offset deleterious effects on accuracy and knowledge-bases. We anticipate these adjustments by aiming to preempt individual and scaled effects through the advocacy of the application of conscious epistemic norms by agents as well the formal codification of these norms at institutional scales. 

As public institutions, public libraries have a deontic duty to play a role in offsetting the negative effects of AI engagement within the general public. The solutions that we present involve fostering norms that increase the incentives and likelihood for human agents to practice and cultivate reflective knowledge. Before outlining these solutions, we situate epistemic practices within shifting information seeking trends and information retrieval paradigms. We argue that knowledge acquisition, learning, and evidence gathering in the current information ecosystem are inextricable from information seeking practices and information retrieval algorithms. We argue that human-LLM interaction presents a paradigm of search and query where information seeking and retrieval are seamlessly integrated within human-chatbot interaction. 

\section{Information Seeking, Retrieval \& LLMs}

In the foregoing, we will conceptually link information seeking practices, extant and emerging models of information retrieval and institutional epistemology. Practices of information seeking and dominant models of retrieval are constitutive elements of evidence gathering, source evaluation, information dissemination, and institutional epistemology more broadly. 

As we have argued, epistemic standards impose constraints on the transmission of reliable knowledge and the dissemination of information more broadly. In order to ensure that reliable and rationally vetted information channels through academic databases, search engines, and other systems of information storage, the algorithmic processes that retrieve and rank-order information sources need to be made transparent. The recent rise of LLMs and wide-scale adoption of LLM-powered chatbots threatens to undermine the traditional systems of information seeking, storage, and retrieval.The training of proprietary LLMs such as ChatGPT3-4 is opaque and not disclosed to the public. The weights of the training models are equally proprietary. While artificial neural networks (ANNs) pose a general black-box problem of interpreting the information processing in the hidden layers, proprietary AI poses an additional layer of black-boxing: the training data and weights of the model cannot be publicly evaluated. The public must therefore rely on internal corporate standards without being able to evaluate the hyperparametralization of LLMs. Consequently, the emergence of proprietary LLMs as a rival platform of information seeking and retrieval presents a host of epistemic threats to institutional epistemology as well as the individual epistemic standards of human agents. 

In the information sciences, information discovery and retrieval constitute separate though deeply interrelated areas of research. Theories of information seeking and discovery such as Bates’s berrypicking model \cite{bates1989design}, Dervin’s sensemaking model \cite{dervin1998sense, naumer2008sense}, information foraging \cite{pirolli1999information} differ in their approaches—ranging from the dynamic, iterative search process in the berrypicking model, to the sensemaking model’s focus on the continuous construction of meaning, and the information foraging theory’s application of ecological principles to user behavior, where individuals are seen as "foragers" navigating information environments, optimizing their search strategies to gather the most valuable resources while minimizing effort. These theories bifurcate between prioritizing the design of efficient systems and prioritizing user experience. For this reason, the rise of user-experience design seeks to develop intuitive, efficient, and adaptive systems that minimize cognitive load and support users in finding relevant information with minimal effort, aligning with the principle of least resistance. Zipf’s principle of least effort \cite{zipf1951human}predicts that users choose the path of least resistance with respect to their information needs. Since users seek the quickest path in satisfying their information needs, LLM chatbot usage is likely to increasingly dominate the share of query-based retrieval and seeking. This does not necessarily imply the homogenization of information seeking methods, but it does suggest that human-LLM interaction fundamentally changes the landscape of both information seeking and retrieval by consolidating them into chatbot interactions. 

Due to the epistemic weight we place on evidence in our internalist view of epistemic justification, information seeking models play an important causal role in epistemic justifiedness. Rational agents seek information within the context of some task, which is in turn nested within the context of a social process. While information-seeking contexts vary, the paradigmatic knowledge context concerns information foraging for evidence. The search process differs between individuals as attested by the diversity of models, but common to all these models is that information seeking must be understood in the context of instrumental action laden with motivation and rewards toward the pursuit of an end even if that end is exploratory in nature. In most contexts, expedient procurement of the relevant information trumps the reflective analysis of that information. Reasoning through large amounts of information, therefore, cannot be explained without recourse to norm-bound social contexts that sanction this behaviour as a social good. Higher institutions of learning are obvious examples that supply the environment, norms, and analytic tools for reflective comportment toward information. In this respect, we view information seeking as malleable within fixed parameters through the imposition of norms. 

Information retrieval, on the other hand, concerns the mechanisms of obtaining information from a repository through user queries. Information retrieval techniques divide into statistical and semantic models, with Machine Learning (ML) bridging the gap \cite{ruambo2019towards}. Statistical models include probabilistic retrieval, boolean, and vector space models. Semantic models include Latent Semantic Analysis (LSA), Latent Dirichlet Allocation (LDA), word embeddings and knowledge graphs \cite{deerwester1990indexing, blei2003latent}. Machine learning (ML) algorithms can be used to enhance both of these retrieval methods through natural language processing (NLP), traditional ML algorithms like linear regression (LR), and enhancement through artificial neural networks (ANNs) such as large language Models (LLMs) \cite{ruambo2019towards}. 

Traditional vector space models represent terms through sparse vectors equal in length to the number of documents called one-hot encoding. Methods such as term-frequency inverse-document frequency (TF-IDF) rely on distance metrics such as cosine similarity to compute the distance or similarity between document vectors \cite{salton1975vector}. This method of encoding term tokens relies on term frequency and thereby misses relationships based on synonymy and polysemy \cite{blei2003latent}. 

LSA and LDA constitute semantic methods that aim to improve recall and precision in the sparse-vector model. LSA seeks to overcome retrieval problems within sparse encoding cooccurrence methods by using singular-value decomposition (SVD) to construct a latent semantic space from the large document-feature matrix \cite{deerwester1990indexing}. LDA, on the other hand, uses a Bayesian hierarchical generative model to generate topics from a corpus of documents \cite{blei2003latent}. LDA takes a Dirichlet distribution as the prior and a corpus of documents as data in order to update its posterior distribution of topic predictions \cite{blei2003latent}. A Dirichlet distribution refers to continuous multivariate distribution that encodes a probability vector over k topics simultaneously \cite{wong1998generalized}. The Dirchlet prior is set manually as a hyperparameter that encodes assumptions about topic sparsity and concentration in order to generate topic distributions per document and word distributions per topic. The topics are defined as distributions over the entire vocabulary of the corpus \cite{blei2003latent}. 

While LSA and LDA constitute important improvements upon vector-space models, the innovation of dense retrieval methods enabled by word embeddings and Transformer architectures have ushered a new paradigm of information retrieval where both queries and documents are encoded into dense-vector representations that capture both local and global semantic relationships. 

Recognizing the prospects of large-scale adoption of LLMs and the imminent ubiquity of Generative AI, IR researchers have proposed Generative Retrieval (GR) models that leverage Generative AI and LLM capabilities to enhance retrieval. Some of these proposals respond directly to the retrieval problems and epistemological challenges posed by LLMs that we have broached. Generative Information Retrieval (GIR) models bifurcate into two approaches: a) Generative Document Retrieval (GDR) and b) Reliable Response Generation (RRG) \cite{li2024matching}. GDR models seek to enhance extant document retrieval methods with generative AI and dense retrieval methods. Standard GDR models train generative models to map queries to relevant DocIDs using sequence-to-sequence (seq2seq) methods. Proposed document Identifier (DocID) methods range from numeric-based to text-based identifiers that can be either static or dynamic. Furthermore, hybrids of generative and dense retrieval methods have been proposed that match queries to document clusters and subsequently perform robust searches that optimize for precision and recall \cite{li2024matching}. 

Different from GDP models, RRG approaches closely mirror the direct information access enabled by LLM-based chatbots. RRG methods seek to optimize internal model representation and update mechanisms through optimizing training data on the one hand and optimizing training methods on the other. Retrieval Augmentation Methods (RAG) include the augmentation of RRG with traditional retrieval mechanisms that fetch relevant documents that can be processed and looped into the response. Challenges that beset RRG approaches include the familiar problems we have discussed surrounding LLMs such as hallucinations, out of date knowledge, lack of context-richness with respect to domain-specific knowledge, bias and information errors \cite{huang2024survey, hersh2024search}. Broadly speaking, the reliability of LLM information retrieval is beset by biased and out of date training data, hallucinatory answers, errors of reasoning, errors of calculation, and lack of domain-specific nuance.

Despite suggesting important improvements, these models fail to recognize the qualitative shift from traditional information retrieval to natural language discursive interactions with LLMs. As I have argued, human-LLM interaction signals a shift not only in information retrieval but also information seeking. The resulting amalgam combines and homogenizes these two processes within a unified interaction model. While querying an LLM bears structural similarities to search in the same way that all cognitive-adjacent tasks involve a system of memory and retrieval, they differ from search in that they structurally resemble dialogical communication and information-transfer akin to human conversation. The upshot is that LLMs constitute discursive interlocutors equipped with general knowledge (barring domain-expert LLMs) and computational capacities that vastly exceed those of individual humans. In other words, the pooling of human knowledge within LLMs (assuming idealized information retrieval and accuracy) lowers incentives for alternative information seeking paths as well as relevant information bearing sources that require additional effort to discover the semantic content of relevance. While GR models aim to improve the retrieval mechanisms through hybrids of traditional and dense-vector representations, it is reasonable to assume that incremental improvements such as those suggested above will approximate a more accurate discursive interlocutor. 

Therefore, these GR solutions are chiefly concerned with improving the reliability of LLMs within the scope of information retrieval but do not address the broader epistemological issues we have raised with respect to collective intelligence, collective epistemology, and collective rationality. In order to redress the epistemological issues we have raised, models of human-AI interaction that aim to offset deleterious effects on reasoning need to be developed. These models should aim to stem the scaled effects of outsourcing reasoning to Generative AI by either conserving or enhancing the epistemic practices of individual human reasoners within human-LLM interaction on the one hand, and improving model performance on the other. The former is a problem concerning norms and the latter is a problem concerning model parameters. These axes result in a trade-off that I will call the AI-performance-reasoning trade-off. The performance reasoning trade-off asserts that as model performance improves, the incentive to reason through tasks diminishes (though factors such as use cases and user personalities are likely to vary). At the local scale of individual human-LLM interaction this trade-off produces an efficiency gain with respect to retrieval but a loss with respect to human epistemic status and virtues (where virtues causally predict epistemic justifiedness). However, this efficiency gain in individual interactions incurs losses to collective intelligence when scaled to collective LLM-retrieval adoption. 

\section{Minimizing Epistemic Threats}

In order to minimize epistemic threats, we propose an interaction model in which human agents exercise informed and strategic agency. To this end, we contrast unplanned and impulsive interaction with interaction that internalizes a strategic model designed to produce optimal outcomes. The strategic model is meant to serve as a flexible blueprint for interactions that avoid long-term epistemic threats and explores pathways that enhance epistemic virtues. Knowledge and education about LLMs constitutes a key component of developing such a model, but it does not alone suffice to seed virtuous interaction habits. For that, prior knowledge about model capabilities need to be wedded to action pathways that veer away from the temptation of instant gratification. To this end, we propose an interaction model that incorporates self-control and delayed gratification theory and normative epistemology within a broader human-LLM interaction model. Our model integrates virtuous information seeking with virtue epistemology.

Virtue epistemologists identify open-mindedness, intellectual courage, intellectual integrity, intellectual humility, and epistemic responsibility as paradigmatic virtues that can shape epistemic norms \cite{choo2016epistemic}. Conversely, they identify gullibility, dogmatism, prejudice, closed-mindedness, and negligence as epistemic vices \cite{cassam2016vice}. While these traits are not meant to be exhaustive, they identify core traits whose cultivation or avoidance produce epistemically optimal outcomes. In our model, the motivation for knowledge constraints the information seeking journey. However, we recognize that agents are not always motivated by knowledge but action outcomes e.g. test answers, assignment completion, letter writing, idea generation. In the ideal scenario, obtaining the requisite knowledge is a necessary step in achieving the outcome. With an LLM in the loop, human agents can substitute the knowledge motivation by asking the LLM to produce the outcome. Virtue epistemology prioritizes the knowledge motivation as a means of achieving the outcome. As we have already argued, this is desirable because human agents possess the capability for reflective knowledge. However, reflective knowledge cannot be acquired without instantiating epistemic virtues.

However, the individual interaction model does not suffice in steering collective action toward minimizing epistemic vices. At the meso-scale of interaction, we propose the creation of broadly agreed-upon norms that negatively sanction harmful interactions and positively sanction beneficial interactions. At this solution scale, public institutions such as libraries have a major role to play in developing guidelines that spread awareness and instill an ethos of positive and negative outcomes within human-LLM interaction.
At the third solution scale, we propose the development of deontic rules that limit harmful interactions with LLMs. We propose three methods of deontic rules: a) legislation that requires LLM models to exhibit discursive norms that steer human agents toward beneficial interactions, b) organizational rules that prohibit harmful uses and c) equipping models with hard-coded rules or discursive norms that steer human interactions towards beneficial interactions.

We propose three types of interrelated and mutually inclusive methods for minimizing epistemic threats: \\

\usetikzlibrary{arrows.meta, shadows, positioning}

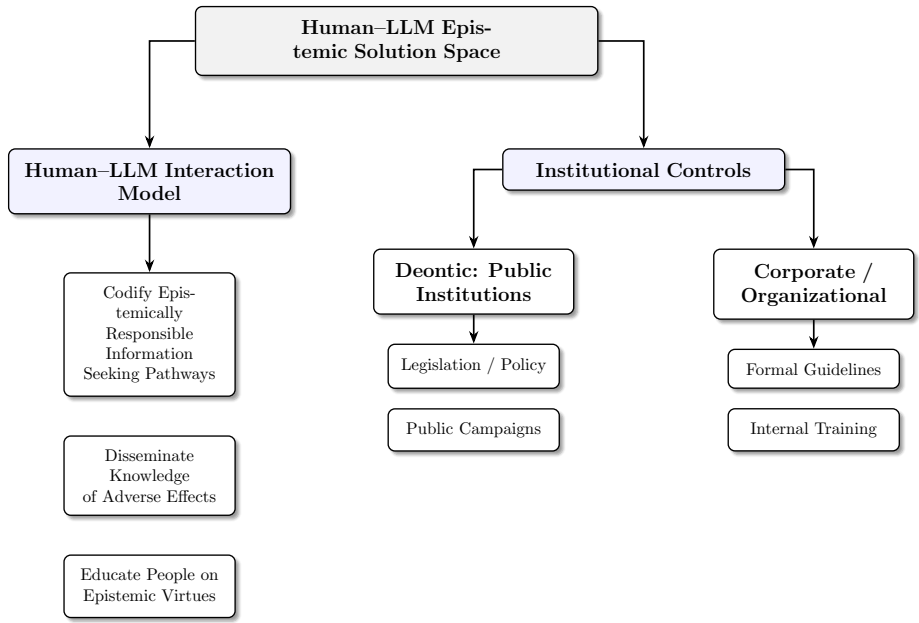
\begin{figure*}[t]
\centering
\resizebox{\textwidth}{!}{%
\begin{tikzpicture}[
    node distance = 1.2cm and 0.5cm,
    every node/.style = {
        draw, rounded corners, 
        fill=white, blur shadow, 
        align=center, font=\large, 
        inner ysep=8pt, inner xsep=5pt,
        line width=0.8pt
    },
    root/.style = {text width=8cm, fill=gray!10},
    main/.style = {text width=5.5cm, fill=blue!5},
    sub/.style = {text width=3.8cm},
    leaf/.style = {text width=3.2cm, font=\normalsize}
]

\node [root] (root) {\bfseries Human--LLM Epistemic Solution Space};

\node [main, below left = 1.5cm and -2cm of root] (interaction) {\bfseries Human--LLM Interaction\\Model};
\node [main, below right = 1.5cm and -2cm of root] (institutional) {\bfseries Institutional Controls};

\node [leaf, below = 1.2cm of interaction] (pathways) {Codify Epistemically\\Responsible Information\\Seeking Pathways};
\node [leaf, below = 0.8cm of pathways] (adverse) {Disseminate Knowledge\\of Adverse Effects};
\node [leaf, below = 0.8cm of adverse] (virtues) {Educate People on\\Epistemic Virtues};

\node [sub, below left = 1.2cm and -1.5cm of institutional] (deontic) {\bfseries Deontic: Public\\Institutions};
\node [sub, below right = 1.2cm and -1.5cm of institutional] (corporate) {\bfseries Corporate /\\Organizational};

\node [leaf, below = 0.6cm of deontic] (leg) {Legislation / Policy};
\node [leaf, below = 0.4cm of leg] (campaigns) {Public Campaigns};

\node [leaf, below = 0.6cm of corporate] (guide) {Formal Guidelines};
\node [leaf, below = 0.4cm of guide] (training) {Internal Training};

\begin{scope}[-Stealth, line width=1pt]
    \draw (root) -| (interaction);
    \draw (root) -| (institutional);
    \draw (interaction) -- (pathways);
    \draw (institutional) -| (deontic);
    \draw (institutional) -| (corporate);
    \draw (deontic) -- (leg);
    \draw (corporate) -- (guide);
\end{scope}

\end{tikzpicture}
}
\caption{Human--LLM Epistemic Solution Space (Vertical TikZ Layout).}
\label{fig:human-llm-solution-space}
\end{figure*}

\paragraph{Epistemological Model of Human-LLM Interaction}
The interaction model involves informing human users of what an LLM is, the harmful effects of misuse and the beneficial effects of proper use. This control aims to provide factual knowledge as well as normative guidelines for avoiding harmful effects and gaining beneficial ones. With this control in place, rational agents are ultimately responsible for delaying instant reward when they interact with an LLM. 

\begin{figure}[t]
  \centering
  \includegraphics[width=\linewidth]{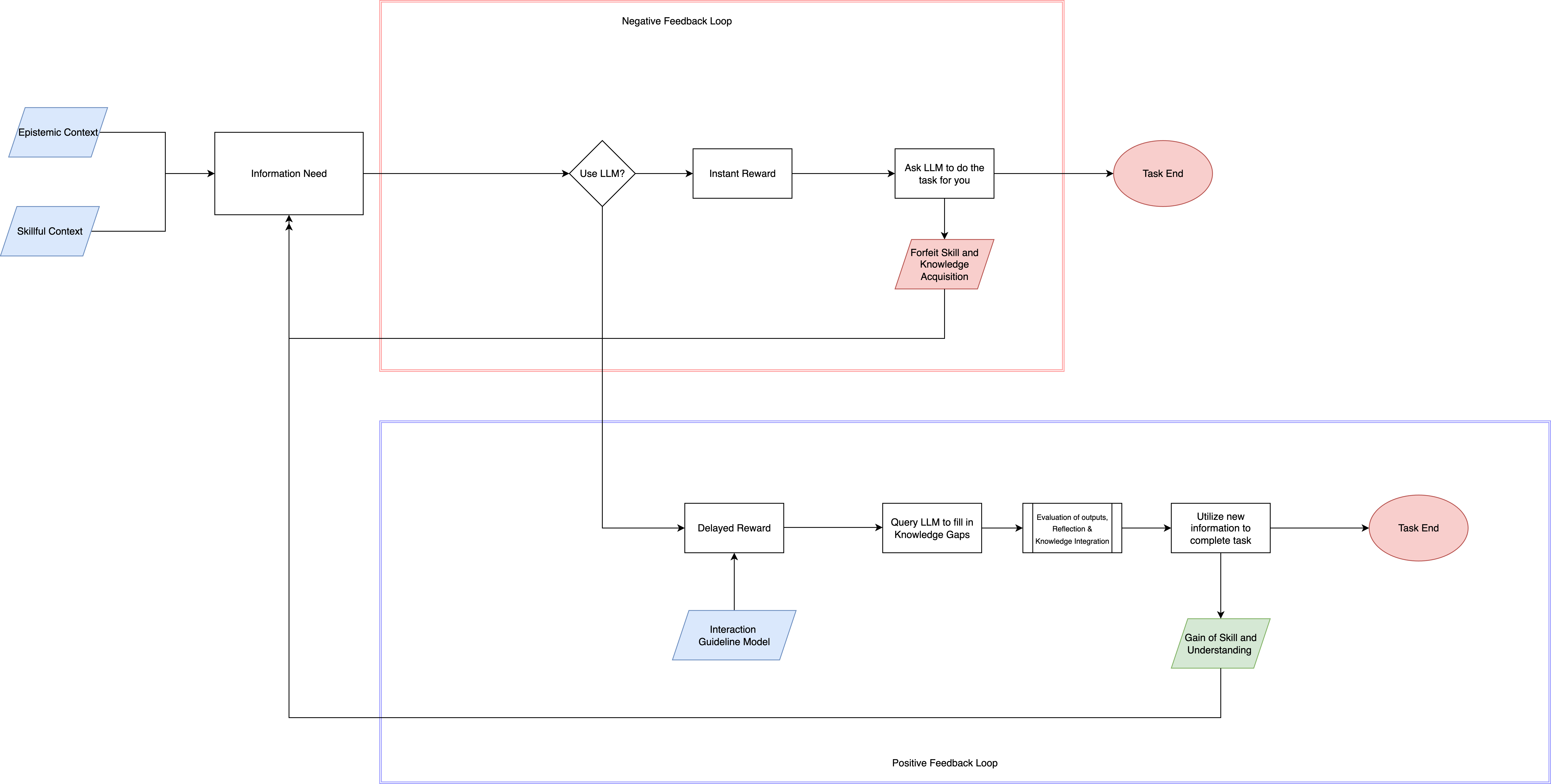}
  \caption{Epistemological Model of Human--LLM Interaction.}
  \label{fig:epistemological-model}
\end{figure}

\paragraph{Human-LLM Interaction Institutional Norms Setting }
The institutional scale of norm setting is meant as a broad campaign from public institutions and internally by private organizations to set collective norms by informing agents . Evidence shows that human agents have a strong norm-conforming disposition (Heath 2011).  Developing an ethos of “dos and don’ts” at the collective level as well as discursive negative sanctioning of unvirtuous uses and positive sanctioning of virtuous uses can mitigate self-control limitations in individual actors. Creating a discourse of potential harms can make individuals cautious about impulsive uses of LLMs. 

\paragraph{Discursive Norms and Deontic Constraint}
This solution approach suggests equipping LLMs with discursive norms that steer users from epistemically harmful uses such as forfeiting knowledge gain and skill utilization for instant information outcomes. A more controversial facet of this solution approach involves hard-coding “dos and don’t” into the model. Possible pathways for this include Constitutional AI (Bai et al. 2022), as well as legislation and public policy. While the deployment of legislation should be approached cautiously, laws and regulations can help limit the epistemic and broader harms of LLMs by restricting harmful usage in more sensitive contexts such as educational institutions. \\

While the adoption and use-cases of LLMs are still in their infancy, we believe that we have identified important ways in which LLMs fundamentally change the information landscape and their epistemological effects on individual rational agents and collective rationality. 

\section{Maximizing Epistemic Virtues}

Maximizing epistemic virtues constitutes the converse of minimizing epistemic threats. In order to stave off the epistemic effects incurred by the principle of least effort, we have devised guidelines aimed at promoting epistemic norms that counteract the potential causal impacts of LLMs. Just as we have outlined the potential negative impacts of LLMs on reasoning, correct usage of LLMs can conversely result in positive impacts. Positive usage of LLMs carries the potential to revolutionize learning provided that human agents prioritize knowledge and skill acquisition over results. With the correct approach, LLMs can enhance learning, understanding, information gathering, and . As we have argued, the greatest advantage that LLMs possess comes from the conjunction of their discursive ability and the  breadth of their information retrieval and synthesis. This means that human agents can query them at a fine-grained linguistic level in order to gain knowledge and understanding of difficult topics, and, provided outstanding problems with hallucinations and outdated knowledge are mitigated, ascertain up-to-date facts and evidence. \\

We argue that cultivated and norm-regulated usage of LLM chatbots can bring the following epistemic advantages to human agents: 
\begin{itemize}
   \item Accurate information retrieval 
   \item Expedient relevance discovery
   \item Gain of understanding
   \item Improved reasoning
   \item Increase of curiosity
   \item Increase of creativity 
   \item Reinforcement of epistemic virtues 
\end{itemize}

\section{Conclusion}

The public availability of LLMs stands to change the relationship that human agents have to knowledge acquisition and information seeking. In this paper we have identified threats that stem from the current state of affairs as well as extrapolated wide-scale effects to collective rationality if these threats are not mitigated. We have argued that the threats are twofold: individual and collective. At the individual level, human-LLM interaction threatens to erode epistemic virtues of knowledge motivation, comprehension and reflective reasoning. At the collective scale, the substitution of LLMs for information seeking, discovery, and task completion could introduce pernicious factual and reasoning errors into chains of institutional decision-making. We have argued that a three tiered-approach to mitigating these effects is desirable: a) the development of human-LLM interaction models with the promotion of epistemic virtues in mind b) the development of institutional and organizational guidelines that steer norm-formation toward more controlled and beneficial human-LLM interactions, and c) the possibility of imposing deontic constraints on individual models such as hard-coded constitutions or leveraging legislation and policy to generate mitigatory frameworks toward the deployment of LLM models in organizational settings. We have further argued that normatively regulated human-LLM interactions could produce epistemically beneficial outcomes by embedding reliable information retrieval within discursive interactions that enhance learning, comprehension and creativity. At the center of our epistemological framing lies the friction between reliable information transmission and reflective rationality. The former only transmits the truth, whereas the latter sets it on a firm foundation. We have argued that LLMs can only instantiate the former, but unchecked, threaten to erode human reflective epistemic standards that produce knowledge. In order to safeguard the production of knowledge from adverse human-LLM interaction, we have firmly argued for the promotion of epistemic virtues at the individual and organizational scales as a means of steering collective norms toward beneficial human-LLM interaction that subserve societal cooperation.

\printbibliography 
\end{document}